\documentclass[aps,pra,twocolumn,superscriptaddress]{revtex4-2}

\usepackage[utf8]{inputenc}
\usepackage[T1]{fontenc}
\usepackage{amsmath,amssymb,amsfonts,mathtools}
\usepackage{bm}
\usepackage{graphicx}
\usepackage{hyperref}
\usepackage{xcolor}
\usepackage{enumitem}

\hypersetup{colorlinks=true,linkcolor=blue!60!black,
            citecolor=blue!60!black,urlcolor=blue!60!black}

\newcommand{\Sb}{\mathbf{S}}
\newcommand{\Bf}{\mathbf{B}}

\newcommand{\rd}{\,\mathrm{d}}

\begin{document}

\title{Cooperative control of dissipative qubits: steering along the fastest dissipative contraction}

\author{Robert Wei\ss}
\affiliation{Max Planck Institute for the Physics of Complex Systems, N\"othnitzer Stra{\ss}e 38, D-01187 Dresden, Germany}

\author{Sandro Wimberger}
\affiliation{Dipartimento di Scienze Matematiche, Fisiche e Informatiche, Universit\`a  di Parma, Parco Area delle Scienze, 53/A, I-43124 Parma, Italy.}
\affiliation{INFN, Sezione di Milano Bicocca, Gruppo Collegato di Parma, Parco Area delle Scienze 7/A, 43124 Parma, Italy}

\author{David Gu\'ery-Odelin}
\affiliation{Laboratoire Collisions Agrégats Réactivité, IRSAMC,\
          Université de Toulouse, CNRS, UPS, 31062 Toulouse, France}

\date{\today}

\begin{abstract}
In the control of dissipative quantum systems, the slow relaxation modes usually
set the ultimate manipulation timescale. We show that this bottleneck can be
bypassed when dissipation itself is used as the engine of the transfer. For a
qubit with non-unital, anisotropic Bloch relaxation characterized by the ratio
$\kappa=T_1/T_2>1$, coherent control can only rotate the Bloch vector, whereas
the bath alone can change its length. The fastest transfer between two thermal
states therefore consists in steering the Bloch vector, at every instant, along
the direction in which the bath contracts it most rapidly. This direction is
obtained in closed form and lies slightly beyond the equatorial plane. With
strong pulses, the resulting minimum transfer time is a fraction of $T_2$,
instead of several $T_1$ for passive relaxation. With bounded fields, the
Pontryagin maximum principle yields a bang--singular--bang synthesis whose
singular arc is precisely this fastest-contraction direction, and the minimum
time exceeds its strong-pulse limit by a correction inversely proportional to
the field amplitude. We also determine the full time--energy Pareto front:
moderate accelerations require little energy, and the time optimum lies at its
steep, high-energy end. Off-axis targets are treated within the same framework;
their only additional benefit is a shorter final rotation. These results apply
directly to magnetic-resonance spins, nitrogen-vacancy centers and
superconducting qubits, all of which operate deep in the strong-pulse regime.
\end{abstract}

\maketitle

Controlling quantum states in the presence of dissipation is a central
problem in quantum information processing~\cite{NielsenChuang}, quantum
metrology~\cite{GiovannettiScience04}, and quantum
thermodynamics~\cite{Vinjanampathy2016}. Beyond its theoretical interest,
this problem underpins several practical bottlenecks of current quantum
technologies. Reliable qubit initialization to a fiducial state is one
of DiVincenzo's necessary criteria for quantum
computation~\cite{DiVincenzo2000}, and the fast active reset of ancilla and
data qubits is a critical resource for quantum error correction, where
leakage and residual excitations must be evacuated on the timescale of a
single gate operation~\cite{Magnard2018,Zhou2021,Kim2024}. The standard passive strategy relies on thermal relaxation, but its characteristic timescale, the longitudinal relaxation time $T_1$, is set by the environment and cannot be shortened by coherent control alone. $T_1$
determines how long a population imbalance survives in the presence of
the environment and ranges from microseconds in superconducting
qubits~\cite{Krantz2019,Kjaergaard2020} to milliseconds in
nitrogen-vacancy centers~\cite{Doherty2013} and seconds in
NMR~\cite{Abragam1961}. A standard active strategy is, conversely, to compensate for the environment: shortcut-to-adiabaticity protocols~\cite{Torrontegui2013AAMOP,Guery-Odelin2019RMP,Alipour2020Quantum,Guery-Odelin2023,Hatomura2024JPhysB}
and geometric dissipation-compensation methods~\cite{Impens2019SciRep}
use coherent fields to suppress or bypass relaxation while steering the
system toward a target state. A parallel line of work in classical stochastic thermodynamics has developed engineered swift equilibration protocols. In those protocols, a Brownian particle or a
micromechanical oscillator is driven to a new equilibrium state in a
time much shorter than its natural relaxation time by a tailored
time-dependent confinement~\cite{Martinez2016NatPhys,LeCunuder2016APL, Raynal2023, Guery-Odelin2023}.
These approaches are best suited when the environment does not select a
preferred direction in state space.

In many relevant open systems, the situation is different. Thermal
relaxation, amplitude damping, and optical pumping do not only reduce
coherences; they also drive the Bloch vector toward a displaced fixed point
inside the Bloch sphere. Rather than treating this drift only as an error source, dissipation- and
reservoir-engineering approaches use it to prepare, stabilize, or protect
selected quantum states~\cite{Poyatos1996,Verstraete2009,Harrington2022}, turning
controlled couplings to a tailored bath into a resource for state
preparation, stabilization, and autonomous error correction.

Even in the absence of bath engineering, a central question is how to exploit dissipation to drive the system from one thermal equilibrium state to another when an external parameter is varied, and how much speedup over passive relaxation such a strategy can provide.

Time-optimal control of dissipative two-level systems has been developed
extensively within geometric control
theory~\cite{Sugny2007,Bonnard2009,Lapert2010,Mukherjee2013,Lokutsievskiy2024}, where
the bang--singular--bang structure of the optimal trajectories is well
established. Here we revisit this structure for a concrete and practically
important task, the transfer of a qubit between two thermal states, and show that
it has a transparent physical content. The singular arc coincides with the direction along which the bath contracts the Bloch vector most rapidly, so that the transfer is driven by the dissipative drift rather than by the coherent field. This logic is related to the
quantum Mpemba effect~\cite{Ares2025}, in which anomalously fast relaxation is
obtained by preparing an initial state whose overlap with the slowest-decaying
Liouvillian eigenmode is suppressed~\cite{Carollo2021}; such methods have been
used to accelerate qubit resets~\cite{Moroder2026}. In the present scheme, the coherent field plays an analogous role continuously: it maintains the residual mismatch in the coherences, which decay at the fast rate $1/T_2$, rather than in the populations, which relax at the slower rate $1/T_1$. The resulting design rules involve two dimensionless parameters: the anisotropy ratio
$\kappa=T_1/T_2$ and the ratio of the available Rabi frequency to $1/T_2$.

Our analysis rests on an elementary observation. A coherent field rotates the Bloch vector but leaves its length $r=|\Sb|$ unchanged; the transfer from one thermal state to a less polarized one ($s_i\to s_f<s_i$) requires a reduction of this length, which only the bath can provide. The control problem is therefore a navigation problem in the sense of Zermelo: the field sets the \emph{orientation} of the Bloch vector, and the bath sets the rate at which it contracts along that orientation.
We first solve the problem with unbounded pulses (Sec.~\ref{sec:impulsive}). The optimal strategy then consists in orienting the vector along the direction of maximal dissipative contraction, which we obtain in closed form; this yields the absolute time floor of the problem. We then impose a bounded field
(Sec.~\ref{sec:bounded}): the Pontryagin maximum principle shows that the same
direction reappears as a singular arc, and gives the minimum time as a function of
the field amplitude. Section~\ref{sec:pareto} computes the time--energy Pareto
front. Section~\ref{sec:nonaxial} treats targets off the field axis, which are not
fixed points of the bath, and Sec.~\ref{sec:discussion} discusses experimental
platforms. Throughout, an \emph{axial} target is the new thermal equilibrium
$s_f\hat z$, which the bath stabilizes once reached; a \emph{non-axial} target lies
off the axis and must be used or held as soon as it is reached.

\section{Bloch dynamics of open two-level systems}
\label{sec:model}

Consider a two-level system (qubit) coupled to a Markovian environment.
In the interaction picture with respect to the free Hamiltonian, the state
$\rho$ evolves according to the Lindblad master equation~\cite{Lindblad1976,GKS1976}
\begin{equation}
  \dot\rho = -\mathrm{i}[H,\rho]
    + \sum_k \Bigl(R_k\rho R_k^\dagger
      - \tfrac{1}{2}\{R_k^\dagger R_k,\rho\}\Bigr),
  \label{eq:lindblad}
\end{equation}
where $H = \gamma \mathbf{B}\cdot\vec\sigma/2$ is the control Hamiltonian
and the $R_k$ are jump operators describing the coupling to the environment.
Expanding $\rho$ in the Pauli basis,
$\rho = (\mathbf{1}+\mathbf{S}\cdot\vec\sigma)/2$,
a standard calculation (see Appendix~\ref{app:lindblad_bloch} for a
self-contained derivation) maps Eq.~\eqref{eq:lindblad} exactly onto an
affine equation for the Bloch vector $\mathbf{S}\in\mathbb{R}^3$:
\begin{equation}
\dot{\mathbf{S}} = \gamma \mathbf{B}\times\mathbf{S}
- \Lambda\mathbf{S} + \mathbf{d},
\label{eq:bloch_affine}
\end{equation}
where $\Lambda$ is a $3\times 3$ real dissipation matrix and
$\mathbf{d}\in\mathbb{R}^3$ is the affine drift generated by the
non-normal part of the jump operators. Explicitly, writing
$R_k=r_{0k}\mathbf{1}+\mathbf{r}_k\cdot\vec\sigma$, one has
$\mathbf{d}=i\sum_k \mathbf{r}_k\times\mathbf{r}_k^\ast$; the full
expression of $\Lambda$ is given in Appendix~\ref{app:lindblad_bloch}.
Whenever the drift vector does not vanish ($\mathbf{d}\neq 0$), at least one jump operator is non-normal and the dissipation matrix $\Lambda$ is automatically invertible (Appendix~\ref{app:Lambda_invertible}). In that case, $-\Lambda\mathbf{S}+\mathbf{d}$ admits a
unique fixed point $\mathbf{S}^* = \Lambda^{-1}\mathbf{d}$, and
Eq.~\eqref{eq:bloch_affine} takes the equivalent form
\begin{equation}
  \dot{\mathbf{S}} = \gamma \mathbf{B}\times\mathbf{S}
    - \Lambda(\mathbf{S} - \mathbf{S}^*),
  \label{eq:bloch_fixed_point}
\end{equation}
the symmetric part of $\Lambda$ being always positive semidefinite.

For a thermal bath at inverse temperature $\beta$
with jump operators $R_- = \sqrt{\Gamma_-}\sigma_-$ and $R_+ = \sqrt{\Gamma_+}\sigma_+$
satisfying the thermal equilibrium condition $\Gamma_-/\Gamma_+ = \mathrm{e}^{\beta\hbar\omega_0}$,
one finds $\Lambda = \mathrm{diag}(\Gamma_\perp,\Gamma_\perp,\Gamma_\parallel)$ with
$\Gamma_\perp = (\Gamma_-+\Gamma_+)/2$ and $\Gamma_\parallel = \Gamma_-+\Gamma_+$,
and the attractor reduces to $\mathbf{S}^* = s_*\hat{z}$ with
$s_* = \tanh\!\bigl(\beta\hbar\omega_0/2\bigr)$. Strict $T_1$ relaxation
alone produces the Redfield equality $\Gamma_\perp = \Gamma_\parallel/2$,
i.e.\ $\kappa = T_1/T_2 = 1/2$. In practice, however, additional
pure-dephasing channels---spin-bath fluctuations in NV centers, charge
or flux noise in superconducting qubits, chemical-shift anisotropy and
spin diffusion in NMR---are ubiquitous and dominate the transverse
decay, raising $\Gamma_\perp$ well above $\Gamma_\parallel/2$ while
leaving $\mathbf{S}^*$ and the longitudinal rate $\Gamma_\parallel$
unchanged. The resulting anisotropic Bloch model with $\kappa > 1$
is therefore the standard phenomenological framework~\cite{Abragam1961,Krantz2019,Doherty2013}
in which experimental relaxation data are reported across these
platforms. Throughout this work, $T_2 = 1/\Gamma_\perp$
denotes the genuine irreversible Markovian transverse decay time (see Appendix~\ref{app:lindblad_bloch}). The phenomenological Bloch equation
then reads
\begin{equation}
  \dot{\mathbf{S}} = \gamma\mathbf{B}\times\mathbf{S}
    - \Gamma_\perp\mathbf{S}_\perp
    - \Gamma_\parallel(S_z - s_*)\hat{z},
  \label{eq:bloch_thermal}
\end{equation}
with $S_z = \mathbf{S}\cdot\hat{z}$ the longitudinal component and
$\mathbf{S}_\perp = \mathbf{S} - S_z\hat{z}$ the transverse part.

The drift vector $\mathbf{d}$ encodes the non-unital part of the dissipation. For a single jump operator $R=r_0\mathbf{1}+\mathbf{r}\cdot\vec\sigma$, $\mathbf{d}=i\mathbf{r}\times\mathbf{r}^\ast$ and it vanishes if and only if $[R,R^\dagger]=0$. With several jump operators, only the total drift $\mathbf{d}=i\sum_k\mathbf{r}_k\times\mathbf{r}_k^\ast$ matters, so unital dynamics may result from cancellations between non-normal channels. This yields a natural dichotomy.
 In the unital case, the affine
drift vanishes, $\mathbf{d}=0$, so that (for invertible $\Lambda$) the
fixed point is $\mathbf{S}^*=0$, i.e.\ the maximally mixed state, and
the environment selects no preferred direction in Bloch space. Pure
dephasing and Pauli channels belong to this class, and the
dissipation-compensation strategy of
Ref.~\cite{Impens2019SciRep} is well suited to it. By contrast, when
$\mathbf{d}\neq 0$ (non-unital dynamics), the constant forcing
displaces the attractor to $\mathbf{S}^*\neq 0$ and the environment
actively biases the motion toward it; this inhomogeneous drift cannot be
removed by a static compensating field~\cite{BreuerPetruccione}.
Thermal relaxation, amplitude damping, and optical pumping all belong
to this class. In the present work, this displaced attractor
$\mathbf{S}^*$ provides the bias used by the cooperative protocol.

Equation~\eqref{eq:bloch_fixed_point} also has a simple Zermelo-like
interpretation~\cite{Zermelo1931,Russell2014}. The dissipative term
$\mathbf{w}(\mathbf{S})=-\Lambda(\mathbf{S}-\mathbf{S}^*)$ plays the role
of a drift field in the Bloch sphere, whereas the coherent term
$\gamma\mathbf{B}\times\mathbf{S}$ rotates the state at fixed norm. The transfer time thus depends on the trajectory direction: relaxation accelerates motion toward $\mathbf{S}^*$ and impedes the reverse.

\section{Transfer problem and impulsive limit}
\label{sec:impulsive}
\label{sec:prerot}

We consider the following protocol scenario. A qubit is initially
prepared at thermal equilibrium with the bath under a static field
$\mathbf{B}_i = B_i\hat{z}$, with longitudinal polarization
$\mathbf{S}_0 = s_i\hat{z}$ where
$s_i = \tanh(\beta\hbar\omega_i/2)$ and $\omega_i = \gamma B_i$.
At $t = 0$ the static field is switched to a new value
$\mathbf{B}_f = B_f\hat{z}$ ($B_f < B_i$), so that the bath-selected
equilibrium is shifted to $\mathbf{S}^* = s_f\hat{z}$ with
$s_f = \tanh(\beta\hbar\omega_f/2) < s_i$, i.e., the length $s_*$ of Sec.~\ref{sec:model} evaluated at $\omega_0=\omega_f$. From now on, the attractor length is denoted by $s_f$ only. Throughout we consider this cooling
quench, $s_i>s_f$; the heating case follows from straightforward sign changes. We
assume the switching to be fast on the dissipative scale
$\Gamma_\parallel^{-1}$ and slow on the bath correlation scale, so that the
Bloch rates $\Gamma_\perp$, $\Gamma_\parallel$ remain unchanged and only the
attractor moves. The task is to drive $\mathbf{S}(t)$ from $\mathbf{S}_0$ to a
target $\mathbf{S}_f$ as rapidly as possible by superposing a transverse control
field $\mathbf{B}_\perp(t)$ on $\mathbf{B}_f$.

When the target is the new attractor, $\mathbf{S}_f = s_f\hat{z}$ (axial
target), the transfer realizes an engineered swift thermalization of the qubit
between two thermal equilibria, and the bath itself stabilizes the target once it
is reached. When the target lies off axis,
$\mathbf{S}_f = s_f(\sin\theta_f,0,\cos\theta_f)$ with $\theta_f > 0$
(non-axial target), it is not a bath equilibrium; it is a finite-time transfer
objective, treated in Sec.~\ref{sec:nonaxial}. The passive reference protocol
below is complete when $|\mathbf{S}(t)-\mathbf{S}_f| < \varepsilon$ for a fixed
tolerance $\varepsilon > 0$.

Numbers quoted below refer to the reference quench $\beta\hbar\omega_i=2$,
$\beta\hbar\omega_f=1.2$, i.e., $s_i=\tanh1\simeq0.762$ and
$s_f=\tanh0.6\simeq0.537$, with $\varepsilon=5\times10^{-3}$, and times are
given in units of $\Gamma_\parallel^{-1}=T_1$.

\paragraph*{Passive reference.}
In the \emph{step protocol}, the field is switched from $\Bf_i$ to $\Bf_f$ at
$t=0$ and no transverse field is applied. The Bloch vector stays on the axis and
its longitudinal mismatch decays as $(s_i-s_f)\,e^{-\Gamma_\parallel t}$. The
target is therefore reached within the tolerance $\varepsilon$ after
\begin{equation}
  T_{\rm step}=\Gamma_\parallel^{-1}\ln\frac{s_i-s_f}{\varepsilon},
  \label{eq:Tstep}
\end{equation}
i.e., after several $T_1$. All speedups reported below are defined with respect to this time.

\paragraph*{Radius and orientation.}
We write $\Sb=r\,\mathbf n$ with $r=|\Sb|$ and $|\mathbf n|=1$. Because the
coherent term $\gamma\Bf\times\Sb$ is orthogonal to $\Sb$, it leaves $r$
unchanged: the field acts only on the orientation $\mathbf n$. The radius obeys
$\dot r=\mathbf n\cdot\mathbf w(r\mathbf n)$, where
$\mathbf w=-\Lambda(\Sb-\Sb^*)$ is the dissipative drift. For the thermal bath
Eq.~\eqref{eq:bloch_thermal}, in the meridional plane that contains the axis and
the target ($\Sb=r(\sin\theta,0,\cos\theta)$), this gives
\begin{align}
  \dot r &= b(r,\theta)
  \equiv -r\bigl(\Gamma_\perp\sin^2\theta+\Gamma_\parallel\cos^2\theta\bigr)
          +\Gamma_\parallel s_f\cos\theta,
  \label{eq:rdot_main}\\
  \dot\theta &= a(r,\theta)+u,
  \nonumber\\
  a(r,\theta)&\equiv(\Gamma_\parallel-\Gamma_\perp)\sin\theta\cos\theta
          -\Gamma_\parallel\frac{s_f}{r}\sin\theta,
  \label{eq:thetadot_main}
\end{align}
where $u=\gamma B_\perp$ is the transverse control and $|u|\le u_{\max}$. Excursions out of the meridional plane bring no advantage: since the dissipator does not depend on the azimuth, they merely divert part of the control budget from $\theta$ (Appendix~\ref{app:meridian_proof}). Equation~\eqref{eq:rdot_main}
captures the essential physics of the problem: cooling the qubit from $s_i$ to $s_f$ requires the radius to decrease by $s_i-s_f$, and the polar angle $\theta$ is the only control parameter that sets the rate of this decrease.

\paragraph*{A first strategy: the single tilt.}
The simplest way to exploit $\Gamma_\perp>\Gamma_\parallel$ is to rotate the
vector instantaneously by the angle $\psi^*=\arccos(s_f/s_i)$, which places it at
the target latitude ($S_z=s_f$), and then to let it relax freely. The whole remaining mismatch, $\sqrt{s_i^2-s_f^2}$, is then purely transverse, i.e., stored in coherences, which the bath erases at the fast rate $\Gamma_\perp$, and the target is reached
after
$t^*_{\rm coop}=\Gamma_\perp^{-1}\ln(\sqrt{s_i^2-s_f^2}/\varepsilon)$. The gain
over the step protocol is therefore $T_{\rm step}/t^*_{\rm coop}=\kappa\xi$, with
$\xi=\ln[(s_i-s_f)/\varepsilon]/\ln[\sqrt{s_i^2-s_f^2}/\varepsilon]\lesssim1$
($\kappa\xi=4.1$ for $\kappa=5$). This strategy captures the correct intuition, namely that coherences decay faster than populations relax, but it is not optimal: after the tilt, the transverse component must decay \emph{below the tolerance}, which adds a $\ln(1/\varepsilon)$ contribution to the transfer time. Since rotations cost no time in the impulsive limit, it is preferable to contract the vector first and to rotate it onto the target at the end.

\paragraph*{Steering along the fastest contraction.}
With instantaneous rotations, $\theta$ can be chosen freely at every instant, and $r$ is the only remaining dynamical variable. The fastest way to bring $r$ from $s_i$ to $s_f$ is therefore to maximize the contraction rate $-\dot r$ at each radius. Setting $\partial_\theta b=0$ in Eq.~\eqref{eq:rdot_main} gives the
\emph{direction of fastest contraction},
\begin{equation}
  \cos\theta_s(r)=-\frac{\Gamma_\parallel s_f}{2(\Gamma_\perp-\Gamma_\parallel)\,r}
  =-\frac{s_f}{2(\kappa-1)\,r},
  \label{eq:theta_s}
\end{equation}
which exists for all $r\ge s_f$ as soon as $\kappa\ge3/2$. This threshold
has a simple meaning. Besides $\theta_s$, the antiparallel orientation
$\theta=\pi$ also satisfies $\partial_\theta b=0$. There the vector points away
from the attractor, so that longitudinal relaxation and the pull of the bath add
up and the contraction rate is $-\dot r=\Gamma_\parallel(r+s_f)$. Tilting the
vector away from $\theta=\pi$ gains the excess transverse rate
$(\Gamma_\perp-\Gamma_\parallel)r$ but loses part of the pull
$\Gamma_\parallel s_f$, and it pays off only when $r>s_f/[2(\kappa-1)]$. Since
the transfer ends at $r=s_f$, the condition $\kappa\ge3/2$ guarantees that the
anisotropy outweighs the pull of the bath all the way to the target, so that
$\theta=\pi$ is a contraction minimum and plays no role. Below this threshold,
the fastest direction would reach $\theta=\pi$ before the end of the transfer,
i.e., the protocol would pass through a population-inverted state. All the
platforms of Table~\ref{tab:feasibility_v19} lie well above it. Two remarks clarify the physical content of Eq.~\eqref{eq:theta_s}. (i) For $\kappa\gg1$ one has $\theta_s\to\pi/2$:
the vector lies in the transverse plane, where its whole length is carried by
coherences that decay at $\Gamma_\perp$. (ii) For finite $\kappa$ the optimum lies
slightly \emph{beyond} the equator ($\theta_s=95^\circ$--$97^\circ$ for
$\kappa=5$, $111^\circ$--$120^\circ$ for $\kappa=2$). There the source term
$\Gamma_\parallel s_f \cos\theta$, i.e., the pull of the bath toward its fixed
point $s_f\hat z$, points inward and adds to the contraction instead of opposing
it. Along $\theta_s(r)$ the contraction law becomes linear in $r^2$,
\begin{equation}
  \frac{\rd r^2}{\rd t}=-2\Gamma_\perp\bigl(r^2+\rho^2\bigr),
  \qquad
  \rho^2=\frac{s_f^2}{4\kappa(\kappa-1)},
  \label{eq:r2_law}
\end{equation}
which integrates in closed form. Including an instantaneous rotation at each end
(onto $\theta_s$ at $t=0^+$, onto the target at the end) yields the minimum transfer time with unbounded control:
\begin{equation}
  t_{\rm floor}=\frac{1}{2\Gamma_\perp}
  \ln\frac{s_i^2+\rho^2}{s_f^2+\rho^2}
  \;\xrightarrow{\;\kappa\gg1\;}\;
  T_2\ln\frac{s_i}{s_f}.
  \label{eq:tfloor}
\end{equation}
Equation~\eqref{eq:tfloor} is the central result of this section. It is an exact lower bound for the
problem: no protocol, whatever its pulse shape, can be faster, since none can contract the vector at a rate exceeding $-b(r,\theta_s)$. The target is reached
\emph{exactly} in finite time. The arrival tolerance no longer enters, and the transfer duration is a fraction of $T_2$ (for the reference quench,
$\ln(s_i/s_f)=0.35$). The corresponding speedup,
\begin{equation}
  \frac{T_{\rm step}}{t_{\rm floor}}
  =\frac{2\kappa\,\ln[(s_i-s_f)/\varepsilon]}{\ln[(s_i^2+\rho^2)/(s_f^2+\rho^2)]}
  \;\simeq\;\kappa\,\frac{\ln[(s_i-s_f)/\varepsilon]}{\ln(s_i/s_f)},
  \label{eq:Sfloor}
\end{equation}
is linear in $\kappa$. Unlike $\xi$, its prefactor is not bounded by unity, since it reflects the $\ln(1/\varepsilon)$ contribution that affects the passive protocol but not the cooperative one ($55$ for $\kappa=5$; see Table~\ref{tab:axial}).
In this limit the transfer time depends on the target only through its length: any state of radius $s_f$, on or off the axis, is reached in the same time $t_{\rm floor}$, a property used in Sec.~\ref{sec:nonaxial}.

The single tilt and the optimal strategy can be compared in terms of populations
and coherences. The single tilt converts the population excess into coherence once and then relies on free relaxation. The optimal strategy maintains this excess in the form of coherence throughout the transfer, at the angle where the bath erases it most rapidly, and restores the target populations with a final rotation.

\section{Bounded control: the time-optimal synthesis}
\label{sec:bounded}

Physical pulses have a finite amplitude, $|u|\le u_{\max}$, so that rotations now require a finite time. We seek the protocol that reaches the target in minimum time.

\paragraph*{Pontryagin analysis.}
The reduced problem~\eqref{eq:rdot_main}--\eqref{eq:thetadot_main} has the
Pontryagin Hamiltonian $H=p_r\,b+p_\theta(a+u)-1$. The control enters linearly,
so on any interval where $p_\theta\neq0$ the field is saturated,
$u=u_{\max}\,{\rm sign}\,p_\theta$ (a \emph{bang}). The only alternative is a
\emph{singular arc}, on which $p_\theta\equiv0$. Requiring $p_\theta$ to remain
zero gives $\dot p_\theta=-p_r\,\partial_\theta b=0$, and since $p_r\neq0$
(because $H=0$ implies $p_r b=1$), the singular arcs satisfy
\begin{equation}
  \partial_\theta b(r,\theta)=0
  \;\Longleftrightarrow\;
  \theta=0\quad\text{or}\quad\theta=\theta_s(r).
  \label{eq:singular}
\end{equation}
The singular arc is thus the direction of fastest contraction,
Eq.~\eqref{eq:theta_s}. The strategy obtained in the impulsive limit is therefore not an approximation: it is embedded in the structure of the exact bounded problem. The singular arc on the axis ($\theta=0$) corresponds to passive relaxation. The control
that holds the vector on $\theta_s(r)$ is $u_s=\theta_s'(r)\,b-a$. This control is weak, $|u_s|\le1.1\,\Gamma_\parallel$ for $\kappa=5$ and $1.7\,\Gamma_\parallel$ for $\kappa=2$, since it merely compensates the angular drift of the bath. A less obvious consequence is that a free-drift arc with $u=0$ at an angle other than $0$ or $\theta_s$ is neither a bang nor a singular arc, and therefore cannot belong to a time-optimal protocol. In particular, the free-drift arc of the single tilt is
not an extremal, which is why optimizing within that family
($T=0.936\,\Gamma_\parallel^{-1}$ at $\kappa=5$) misses the optimum ($0.150$) by
a factor of six. On the singular arc, the transfer is thus performed by the bath, the weak field serving only to hold the vector along the direction of fastest contraction.

\begin{figure}[t]
\includegraphics[width=\columnwidth]{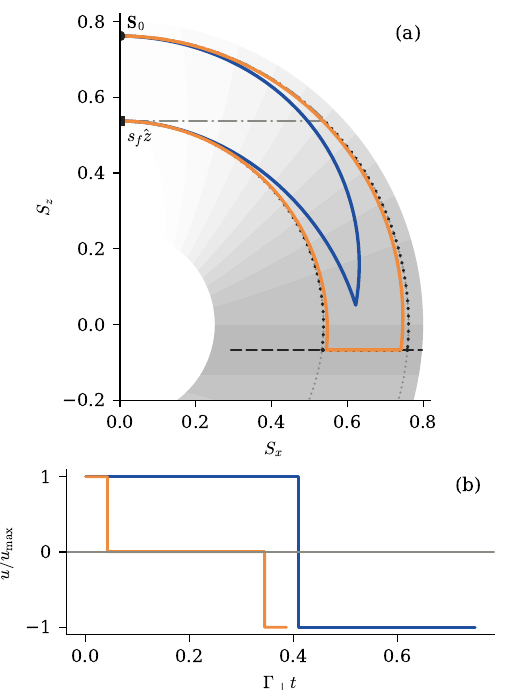}
\caption{Time-optimal transfer to the axial target, $\kappa=5$.
(a)~Meridional plane $(S_x,S_z)$. Circle: initial state $\mathbf S_0=s_i\hat z$;
square: target $s_f\hat z$. Grey shading: dissipative contraction rate
$-\dot r/r$ (darker is faster). Black dashed line: direction of fastest
contraction $\theta_s(r)$, Eq.~\eqref{eq:theta_s}, close to the equator for
$\kappa=5$. Grey dotted arcs: Bloch vectors of length $s_i$ and $s_f$. Grey
dash-dotted line: single tilt to $\psi^*$, followed by the free decay of the
coherences at fixed $S_z=s_f$. Solid blue line: time optimum for
$u_{\max}=20\,\Gamma_\parallel$, a bang--bang that turns back at
$\theta_{\max}=85^\circ$ (Table~\ref{tab:axial}). Solid orange line: time
optimum for $u_{\max}=200\,\Gamma_\parallel$, a bang--singular--bang made of a
rotation along the arc of length $s_i$, a contraction along $\theta_s(r)$ and a
rotation along the arc of length $s_f$; it is almost indistinguishable from the
impulsive protocol (black dotted line). (b)~Corresponding controls $u/u_{\max}$
versus $\Gamma_\perp t$, same colors: bang--bang (blue),
$T_{\min}=0.150\,\Gamma_\parallel^{-1}$; bang--singular--bang (orange),
$T_{\min}=0.077\,\Gamma_\parallel^{-1}$. On the singular arc the field is weak,
$|u_s|\simeq\Gamma_\parallel$, i.e., $|u_s|/u_{\max}\simeq5\times10^{-3}$,
which is invisible on this scale.}
\label{fig:mech_v19}
\end{figure}

The time-optimal protocols therefore consist of bangs and arcs on $\theta_s$, and
only two structures occur (Fig.~\ref{fig:mech_v19}):
1)~\emph{Bang--singular--bang} $(+,s,-)$ for $u_{\max}$ large enough. A first
bang lifts the vector to $\theta_s$, a singular arc follows, along which it contracts at the maximal rate, and a final bang rotates it onto the target. As $u_{\max}\to\infty$ this is
the protocol of Sec.~\ref{sec:impulsive}. 2)~\emph{Bang--bang} $(+,-)$ for moderate $u_{\max}$. The two rotations then last long enough for the contraction accumulated during the bangs to be sufficient, and the vector turns back before reaching $\theta_s$.

In both cases the target is reached exactly. Table~\ref{tab:axial} lists the
minimum times for $u_{\max}=20\,\Gamma_\parallel$. For $\kappa=5$ the optimum is a
bang--bang with turning point $\theta_{\max}=85^\circ$ and
$T_{\min}=0.150\,\Gamma_\parallel^{-1}$, corresponding to a speedup of $25$, six times larger than that of the single tilt. For $\kappa=2$ a singular arc appears at
$\theta_s\simeq111^\circ$--$120^\circ$. These optima agree with unconstrained
direct optimization and with independent 3D GRAPE searches
(Appendix~\ref{app:numerics}).

\begin{figure}[t]
\includegraphics[width=\columnwidth]{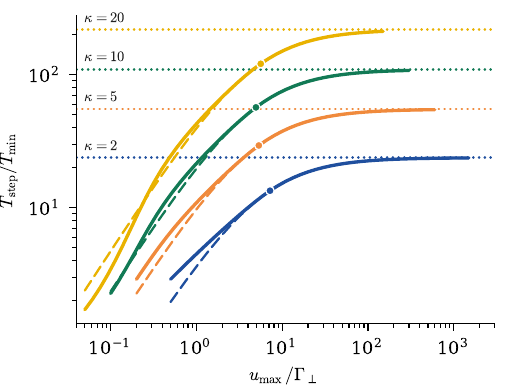}
\caption{Speedup $T_{\rm step}/T_{\min}$ over passive relaxation versus field
amplitude $u_{\max}/\Gamma_\perp$, for $\kappa=2,5,10,20$ (labels; blue, orange,
green, yellow). Solid: exact optimum; dashed:
asymptotic law~\eqref{eq:Tmin_asym}; dotted: impulsive floor~\eqref{eq:Sfloor};
circles: onset of the singular arc. All curves bend at
$u_{\max}\approx4$--$7\,\Gamma_\perp$: below this scale the gain is limited by the
rotation time, above it by the anisotropy.}
\label{fig:speedup_v19}
\end{figure}

\paragraph*{How the field amplitude limits the gain.}
For $u_{\max}\gg\Gamma_\perp$ the bangs are short. Their cost is their duration, $\theta_s/u_{\max}$ each, reduced by the contraction achieved during them. Expanding to first order in $1/u_{\max}$ gives
\begin{equation}
  T_{\min}\simeq t_{\rm floor}+\frac{C(\kappa)}{u_{\max}},
  \qquad
  C=\Phi(s_i)+\Phi(s_f),
  \label{eq:Tmin_asym}
\end{equation}
where each bang contributes the angular penalty
\begin{equation}
  \Phi(r)=\int_0^{\theta_s(r)}\Bigl[1-\frac{b(r,\theta)}{b(r,\theta_s)}\Bigr]\rd\theta,
  \label{eq:Phi}
\end{equation}
with $C\to\pi/2$ for $\kappa\gg1$ ($C=1.79,\,1.61,\,1.59$ for $\kappa=2,5,10$).
Figure~\ref{fig:speedup_v19} shows that Eq.~\eqref{eq:Tmin_asym} is accurate to
four digits throughout the bang--singular--bang regime and remains accurate down to
$u_{\max}\sim\Gamma_\perp$. It also identifies the second characteristic scale of the problem: the rotation time ceases to limit the transfer once
$u_{\max}\gtrsim u^*\equiv C/t_{\rm floor}\simeq\pi\Gamma_\perp/[2\ln(s_i/s_f)]$,
i.e., $u^*\approx4.5\,\Gamma_\perp$ for the reference quench, almost independently
of $\kappa$. The onset of the singular arc occurs at the same scale. Below $u^*$
the gain is limited by the field amplitude, not by the anisotropy. This explains
why the speedup at fixed $u_{\max}=20\,\Gamma_\parallel$ saturates around
$40$--$50$ for large $\kappa$ (Table~\ref{tab:axial}), well below
Eq.~\eqref{eq:Sfloor}. In the opposite limit, $u_{\max}\lesssim\Gamma_\perp$, the angular drift of the bath returns the vector toward the axis faster than the field can tilt it; the gain then drops and $T_{\min}$ grows approximately as $1/u_{\max}$.

\begin{table}[t]
\caption{Axial transfer, reference quench ($s_i=\tanh1$, $s_f=\tanh0.6$,
$\varepsilon=5\times10^{-3}$, $T_{\rm step}=3.80\,\Gamma_\parallel^{-1}$).
$\theta_s$: fastest-contraction angle, Eq.~\eqref{eq:theta_s}, from $r=s_i$ to
$s_f$; $t_{\rm floor}$: impulsive minimum time, Eq.~\eqref{eq:tfloor};
$T_{\min}$, structure and largest polar angle $\theta_{\max}$ reached by
the bang--bang before it turns back (--: bang--singular--bang, which reaches
$\theta_s$): exact optimum at
$u_{\max}=20\,\Gamma_\parallel$ (BB: bang--bang; BSB: bang--singular--bang);
$\kappa\xi$: single-tilt gain. Times in units of $\Gamma_\parallel^{-1}=T_1$.}
\label{tab:axial}
\begin{ruledtabular}
\begin{tabular}{ccccccccc}
$\kappa$ & $\theta_s$ & $t_{\rm floor}$ & $\frac{T_{\rm step}}{t_{\rm floor}}$
 & $T_{\min}$ & synthesis & $\theta_{\max}$ & $\frac{T_{\rm step}}{T_{\min}}$ & $\kappa\xi$\\
\hline
 2 & $111$--$120^\circ$ & 0.160 & 24  & 0.249 & BSB & --           & 15.3 & 1.6\\
 5 & $95$--$97^\circ$   & 0.069 & 55  & 0.150 & BB  & $85^\circ$  & 25.4 & 4.1\\
10 & $92$--$93^\circ$   & 0.035 & 109 & 0.112 & BB  & $62^\circ$  & 34.0 & 8.1\\
20 & $91$--$92^\circ$   & 0.017 & 218 & 0.089 & BB  & $44^\circ$  & 42.7 & 16.3\\
50 & $90$--$91^\circ$   & 0.007 & 545 & 0.079 & BB  & $24^\circ$  & 48.2 & 40.6\\
\end{tabular}
\end{ruledtabular}
\end{table}

\section{Time--energy Pareto front}
\label{sec:pareto}

The time-optimal protocol uses the full amplitude for most of the transfer. Its
pulse energy, $\mathcal E=\tfrac12\int_0^{T}u^2\rd t$, is
$\tfrac12u_{\max}^2T_{\min}$ for a bang--bang ($30\,\Gamma_\parallel$ for
$\kappa=5$, $u_{\max}=20\,\Gamma_\parallel$); the weak singular arc lowers it only
slightly ($40\,\Gamma_\parallel$ instead of $50$ for $\kappa=2$). From a practical standpoint, the relevant question is therefore:
\emph{for a prescribed duration $T$, what is the least energy needed?} Its answer,
$\mathcal E_{\min}(T)$ for $T_{\min}\le T\le T_{\rm step}$, is the Pareto front
of the problem. It connects the passive protocol ($\mathcal E=0$, $T=T_{\rm step}$)
to the time optimum. With a quadratic cost, the maximum principle no longer yields bang arcs, and the optimal field is smooth, $u=\mathrm{sat}_{u_{\max}}(\lambda_xS_z-\lambda_zS_x)$, where $\mathrm{sat}_{u_{\max}}(x)=\max[-u_{\max},\min(u_{\max},x)]$ denotes saturation at $\pm u_{\max}$ and $\bm\lambda=(\lambda_x,\lambda_z)$ is the costate. We solve the resulting boundary-value problem by shooting with
continuation in $T$, keeping the lowest-energy extremal
(Appendix~\ref{app:numerics}).

\begin{figure}[t]
\includegraphics[width=\columnwidth]{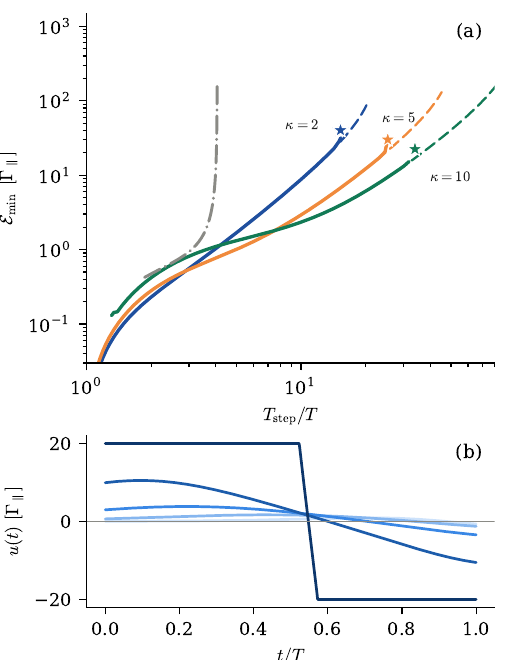}
\caption{(a)~Time--energy Pareto front: minimum pulse energy versus speedup
$T_{\rm step}/T$, for $\kappa=2,5,10$ (exact PMP solutions, same tolerance
$\varepsilon$ as $T_{\rm step}$). Solid: $u_{\max}=20\,\Gamma_\parallel$, ending at
the time optimum (stars: time optimum for the exact target); dashed: unbounded field, which converges to the impulsive floor.
Grey dash-dotted: best single-tilt protocols for $\kappa=5$, which cannot exceed
$\kappa\xi$ and are dominated everywhere. Labels give the value of $\kappa$ (blue: 2, orange: 5, green: 10).
(b)~Optimal controls $u(t)$ versus $t/T$ for $\kappa=5$ and
$u_{\max}=20\,\Gamma_\parallel$, at speedups $T_{\rm step}/T=2,4,8,15$ and $25$
(time optimum), from light to dark blue: smooth tilt-and-return pulses that steepen into the
bang--bang of Fig.~\ref{fig:mech_v19} as $T\to T_{\min}$.}
\label{fig:pareto_v19}
\end{figure}

Figure~\ref{fig:pareto_v19} displays the result. For large $T$ the front decays as
$\mathcal E_{\min}\propto(s_i-s_f)\,e^{-\Gamma_\parallel T}$. Only the residual longitudinal mismatch has to be removed, and the optimal field is active only during the final $\sim T_1$. For $\kappa=5$, the single-tilt gain $\kappa\xi\simeq4$ is
obtained with $\mathcal E\simeq0.8\,\Gamma_\parallel$ and a tilt of only $\simeq18^\circ$,
whereas the single-tilt family needs $\mathcal E\to\infty$ to reach that gain.
A speedup of $10$ requires $\mathcal E\simeq2.9\,\Gamma_\parallel$, one tenth of the energy of the time optimum, and a speedup of $20$ about $12\,\Gamma_\parallel$. The energy diverges
as $T\to T_{\min}$, and the optimal control becomes saturated, ending as the
bang--bang of Sec.~\ref{sec:bounded} [Fig.~\ref{fig:pareto_v19}(b)]. In between, the optimal field retains the same simple shape for all $T$: a smooth
tilt away from the axis followed by a return. As $T$ decreases its amplitude
grows and its turning angle approaches $\theta_s$. Anisotropy turns out to be beneficial only at high speed. A weak field $u$ can only
maintain a small tilt, $S_x\simeq uS_z/\Gamma_\perp$, since the transverse component decays as fast as it is created. The resulting extra
longitudinal relaxation, $uS_x\simeq u^2S_z/\Gamma_\perp$, is \emph{weaker} for
larger $\kappa$. At low energy, anisotropy therefore increases the energy cost instead of reducing it, and the fronts for different $\kappa$ cross
[Fig.~\ref{fig:pareto_v19}(a)]: $\kappa=2$ is the cheapest below a speedup of
$\approx3$, and $\kappa=5$ is cheaper than $\kappa=10$ below $\approx7$. A large $\kappa$ becomes advantageous only at high speedups,
where its front extends much further.

The amplitude bound matters only near $T_{\min}$. Elsewhere the bounded and
unbounded fronts coincide [solid and dashed in Fig.~\ref{fig:pareto_v19}(a)].

\section{Non-axial targets}
\label{sec:nonaxial}

We now consider a target $\Sb_f=s_f(\sin\theta_f,0,\cos\theta_f)$ off the axis. Since it is not a fixed point of the bath, the controlled trajectory must reach it in passing, and the state must then be used immediately (see below). The construction above applies directly, with a single modification: the final bang now stops at
$\theta_f$ instead of returning to the axis.

\begin{figure}[t]
\includegraphics[width=\columnwidth]{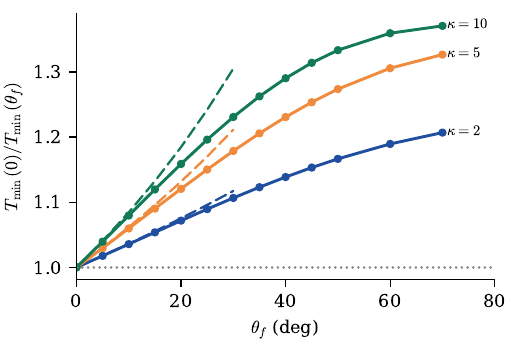}
\caption{Off-axis targets: ratio $T_{\min}(0)/T_{\min}(\theta_f)$ of the axial to
the off-axis minimum time versus the target angle $\theta_f$, for $\kappa=2,5,10$
(labels) and $u_{\max}=20\,\Gamma_\parallel$. Symbols and solid lines: exact
optimum. Dashed: shorter final rotation,
Eq.~\eqref{eq:nonaxial}. Dotted: impulsive limit, in which on- and off-axis
targets of equal length are reached at the same time.}
\label{fig:nonaxial_v19}
\end{figure}

The impulsive limit clarifies the comparison. The time
$t_{\rm floor}$ depends only on the length of the target vector
(Sec.~\ref{sec:impulsive}), so an off-axis target of length $s_f$ is reached
\emph{exactly as fast} as the axial one. Off-axis targets therefore offer no intrinsic geometric advantage. At finite amplitude, they shorten the final rotation,
and for small $\theta_f$
\begin{equation}
  T_{\min}(\theta_f)\simeq T_{\min}(0)-\theta_f/u_{\max},
  \label{eq:nonaxial}
\end{equation}
which is accurate up to $\theta_f\approx20^\circ$ (Fig.~\ref{fig:nonaxial_v19}).
For $\kappa=5$ and $u_{\max}=20\,\Gamma_\parallel$, the target at
$\theta_f=30^\circ$ is reached in $0.127\,\Gamma_\parallel^{-1}$ (gain $30$ over
$T_{\rm step}$), a factor $1.18$ faster than the axial target. The gain grows
slowly with $\theta_f$ and has no upper cutoff: the angle $\theta_f^{\max}$ that
limits interception by free drift plays no role once the final rotation is
admitted. Because the advantage is entirely a rotation-time effect, it vanishes
in the strong-pulse regime $u_{\max}\gg u^*$ in which the experimental platforms
operate.

\paragraph*{Using the intercepted state.}
Since $\mathbf S_f$ is not a fixed point, the protocol is a hitting primitive:
the intercepted state must be used at the hitting time, e.g., read out or
tomographed, or mapped by a fast unitary gate into a decoherence-protected
subspace.

\section{Discussion and conclusion}
\label{sec:discussion}

Non-unital dissipation provides a drift that can be used as a control resource,
but its optimal use differs from the intuitive picture of a single tilt
followed by free relaxation. Coherent control cannot change the length of the
Bloch vector; only the bath can. In the optimal protocol, the field therefore sets the orientation and the bath performs the contraction. The direction is
the fastest-contraction angle $\theta_s(r)$, given by Eq.~\eqref{eq:theta_s},
and it appears in three forms: as the optimal steering law of the impulsive
limit, as the singular arc of the bounded problem, and as the turning point
toward which the minimum-energy pulses converge as $T$ decreases. The resulting
time floor, Eq.~\eqref{eq:tfloor}, is a fraction of $T_2$ and is reached exactly.
At finite amplitude, Eq.~\eqref{eq:Tmin_asym} quantifies the cost of the rotations,
and the Pareto front shows that most of the practical gain is available at
moderate energy.

Two parameters therefore determine whether a given platform can benefit from the mechanism: the anisotropy $\kappa$
and the ratio $u_{\max}/\Gamma_\perp$, which should exceed $u^*/\Gamma_\perp\approx
\pi/[2\ln(s_i/s_f)]$, i.e., a few units. Table~\ref{tab:feasibility_v19} shows
that the three standard platforms satisfy the second condition by orders of
magnitude. A $\pi$ pulse lasts much less than $T_2$, so all three operate in
the regime where Eq.~\eqref{eq:tfloor} applies and the gain is set by $\kappa$.
Only a genuine irreversible Markovian $T_2$ enters, not the inhomogeneous
dephasing time $T_2^*$, which stems from a static spread of precession
frequencies and which a spin echo would simply reverse;
and the hierarchy $\tau_c\ll\tau_p\ll T_2$ guarantees the validity of the
Markovian Bloch description at the shortest pulses.

\begin{table}[t]
\caption{Order-of-magnitude feasibility. $u_{\max}/\Gamma_\perp\approx\pi T_2/\tau_p$
compares the rotation rate with the transverse decay; all platforms lie far
above $u^*/\Gamma_\perp\approx4.5$, so the transfer time approaches
$t_{\rm floor}\approx T_2\ln(s_i/s_f)$ and the gain over passive reset is
$\simeq\kappa\ln[(s_i-s_f)/\varepsilon]/\ln(s_i/s_f)$, Eq.~\eqref{eq:Sfloor}
(last column: reference quench, $\varepsilon=5\times10^{-3}$). The quoted
ranges are indicative, not record values, and depend on sample, operating point,
and on whether the transverse decay is irreversible or refocusable. Transmon
values are guided by Refs.~\cite{Krantz2019,Kjaergaard2020,Place2021,Wang2022,Barends2014};
NV-center values by
Refs.~\cite{Doherty2013,Jarmola2012,Balasubramanian2009,Vallabhapurapu2021};
and NMR values by Refs.~\cite{Abragam1961,Levitt2008,WuOtting2005}.}
\label{tab:feasibility_v19}
\scriptsize
\begin{ruledtabular}
\begin{tabular}{lcccccc}
Platform & $T_1$ ($\mu$s) & $T_2$ ($\mu$s) & $\kappa$ & $\tau_p$ ($\mu$s)
 & $u_{\max}/\Gamma_\perp$ & Gain\\
\hline
Transmon & $50$--$500$ & $2$--$50$ & $5$--$50$ & $0.01$--$0.05$
 & $10^{2}$--$10^{4}$ & $50$--$500$\\
NV center & $10^3$--$10^4$ & $1$--$10^3$ & $10$--$10^3$ & $0.005$--$0.1$
 & $30$--$10^{6}$ & $10^2$--$10^4$\\
NMR & $10^6$--$10^8$ & $10^3$--$10^5$ & $10$--$10^3$ & $5$--$50$
 & $60$--$6\times10^{4}$ & $10^2$--$10^4$\\
\end{tabular}
\end{ruledtabular}
\end{table}

\paragraph*{Robustness.}
The time-optimal protocol is open loop: its switching times are designed for
nominal values of $T_1$ and $T_2$. We have simulated the $\kappa=5$ optimum
($u_{\max}=20\,\Gamma_\parallel$) with $T_1$ and $T_2$ each perturbed
independently by up to $\pm\delta$. The duration is unchanged, and the calibration error results in a deviation from the target, whose worst case grows
linearly with $\delta$: $\simeq2\varepsilon$, $4.4\varepsilon$ and $9.5\varepsilon$
for $\delta=5\%$, $10\%$ and $20\%$ (the same within 3\% at $\kappa=10$). The protocol thus tolerates moderate calibration errors; at the $\pm20\%$ level, a short corrective cycle or feedback is advisable. For the axial target the
residual miss then relaxes toward $s_f\hat z$ under the bath itself.

Several extensions follow naturally. For an arbitrary symmetric dissipation
matrix the impulsive argument applies unchanged: the optimal protocol maximizes
the contraction rate $-\mathbf n\cdot\mathbf w(r\mathbf n)$ over the whole unit
sphere at each radius, and the minimum time is again obtained by a
one-dimensional quadrature (Appendix~\ref{app:general_Lambda}). For $N$-level
systems, the analogue of $r$ is the purity, which unitary control cannot change
either, and the same separation between coherent steering and dissipative
contraction is expected to apply. It also remains to be explored how the mechanism
combines with engineered dissipation~\cite{Verstraete2009,Kraus2008,Witthaut2008}
and with gate design that exploits the non-unital drift.

\paragraph*{Acknowledgments.}
We thank Francesco Petiziol for a critical reading of the manuscript.
D.G.-O. acknowledges support from Institut Universitaire de France and
the ANR project QuCoBEC (ANR-22-CE47-0008). S.W. thanks for funding
from Q-DYNAMO (EU HORIZON-MSCA-2022-SE-01) with project No.
101131418, and from the Italian MUR National Recovery and Resilience
Plan, through the ``National Quantum Science and Technology
Institute'' (NQSTI), spoke~1, project No. PE0000023, CUP
D93C22000940001.

\appendix

\section{From the Lindblad equation to the affine Bloch equation}
\label{app:lindblad_bloch}

We derive the affine Bloch equation~\eqref{eq:bloch_affine} from the
Lindblad master equation~\eqref{eq:lindblad} and establish the explicit
form of $\Lambda$ and $\mathbf{d}$.  

Any $2\times 2$ operator can be expanded in the basis
$\{\mathbf{1},\sigma_1,\sigma_2,\sigma_3\}$.  For the density matrix
and for a single jump operator $R_k$ one writes
\begin{equation}
  \rho = \tfrac{1}{2}(\mathbf{1}+S_i\sigma_i),
  \qquad
  R_k = r_0^{(k)}\mathbf{1} + r_i^{(k)}\sigma_i,
  \label{eq:app_decomp}
\end{equation}
where $S_i\in\mathbb{R}$ and $r_0^{(k)},r_i^{(k)}\in\mathbb{C}$
(summation over repeated spatial indices $i=1,2,3$ is implied).
Using $\dot\rho = \frac{1}{2}\dot S_i\sigma_i$ one immediately identifies
$\dot\rho \;\widehat{=}\; \tfrac{1}{2}\dot{\mathbf{S}}$.

With $H = h_0\mathbf{1}+h_i\sigma_i$ (real $h_i$) the Pauli identity
$[\sigma_i,\sigma_j]=2\mathrm{i}\varepsilon_{ijk}\sigma_k$ gives
$-\mathrm{i}[H,\rho] \;\widehat{=}\; \mathbf{h}\times\mathbf{S}$,
which is a contribution to $\dot\rho$, not directly to $\dot{\mathbf{S}}$.
Since $\dot\rho \widehat{=} \tfrac{1}{2}\dot{\mathbf{S}}$, the corresponding
Bloch-vector equation is $\dot{\mathbf{S}}_{\rm Larmor} = 2\,\mathbf{h}\times\mathbf{S}$.
For $H=\gamma\mathbf{B}\cdot\vec\sigma/2$ one has $h_i = \gamma B_i/2$, so
\begin{equation}
  \dot{\mathbf{S}}_{\rm Larmor}
  = 2\,\mathbf{h}\times\mathbf{S}
  = \gamma\mathbf{B}\times\mathbf{S},
  \label{eq:larmor_check}
\end{equation}
recovering the standard Larmor precession of Eq.~\eqref{eq:bloch_affine}.

For a single jump operator $R = r_0\mathbf{1}+r_i\sigma_i$, the dissipator
$\mathcal{D}[R]\rho = R\rho R^\dagger - \frac{1}{2}\{R^\dagger R,\rho\}$
is evaluated by expanding each product in the Pauli basis and applying
$\sigma_i\sigma_j = \delta_{ij}\mathbf{1}+\mathrm{i}\varepsilon_{ijk}\sigma_k$.
Separating the identity and traceless parts one finds~\cite{BreuerPetruccione}
$\mathcal{D}[R]\rho \;\widehat{=}\; -\Lambda^{(R)}\mathbf{S} + \mathbf{d}^{\,(R)}$,
where the $3\times 3$ real matrix $\Lambda^{(R)}$ and the vector
$\mathbf{d}^{\,(R)}$ are
\begin{align}
  \Lambda^{(R)}_{ij}
    &= |\mathbf{r}|^2\delta_{ij}
      - \tfrac{1}{2}(r_i r_j^* + r_i^* r_j)
      + \tfrac{\mathrm{i}}{2}\varepsilon_{ijk}
        \bigl(r_0 r_k^* - r_0^* r_k\bigr),
  \label{eq:Lambda_full} \\[4pt]
  d^{(R)}_i &= \mathrm{i}\,(\mathbf{r}\times\mathbf{r}^*)_i,
  \label{eq:d_full}
\end{align}
with $\mathbf{r}=(r_1,r_2,r_3)^\top$.  The first two terms in
Eq.~\eqref{eq:Lambda_full} form the \emph{symmetric} part
$M_{ij} = |\mathbf{r}|^2\delta_{ij} - \tfrac{1}{2}(r_i r_j^* + r_i^* r_j)$,
while the third (antisymmetric) term vanishes when $r_0=0$ (traceless
jump operators such as $\sigma_\pm$). 
More generally, the antisymmetric contribution can be reabsorbed into the system Hamiltonian as a Lamb-shift-like correction~\cite{BreuerPetruccione}; throughout this work we assume such terms to be included in $H$, so that $\Lambda$ effectively reduces to its symmetric (real and positive-semidefinite) part.
The drift vector $\mathbf{d}^{\,(R)}=\mathrm{i}(\mathbf{r}\times\mathbf{r}^*)$ vanishes
if and only if $\mathbf{r}\times\mathbf{r}^*=0$, equivalently $[R,R^\dagger]=0$.

Applying Sylvester's criterion to $M$, the three principal minors are
\begin{align}
  P_1 &= |r_2|^2+|r_3|^2 \geq 0, \\
  P_2 &= |\mathbf{r}|^2|r_3|^2 + \tfrac{1}{4}[\mathrm{Im}(r_1 r_2^*)]^2 \geq 0, \\
  P_3 &= \tfrac{|\mathbf{r}|^2}{4}\bigl[
          |\mathrm{Im}(r_1 r_2^*)|^2
         +|\mathrm{Im}(r_1 r_3^*)|^2
         +|\mathrm{Im}(r_2 r_3^*)|^2
        \bigr] \geq 0,
\end{align}
so $M\geq 0$ for any jump operator $R$.  Since the antisymmetric part of
$\Lambda^{(R)}$ does not contribute to
$\mathbf{S}^\top\Lambda^{(R)}\mathbf{S}$, the full matrix is positive
semidefinite in the sense $\mathbf{S}^\top\Lambda^{(R)}\mathbf{S}\geq 0$.

For $N$ jump operators $\{R_k\}$ the contributions add independently:
$\Lambda = \sum_k \Lambda^{(R_k)}$, $\mathbf{d} = \sum_k \mathbf{d}^{\,(R_k)}$.
The symmetric part of $\Lambda$ is a sum of positive-semidefinite matrices,
hence positive semidefinite.
For the thermal model, $R_- = \sqrt{\Gamma_-}\,\sigma_-$ and
$R_+ = \sqrt{\Gamma_+}\,\sigma_+$ (both with $r_0=0$) give
$\Lambda=\mathrm{diag}(\Gamma_\perp,\Gamma_\perp,\Gamma_\parallel)$
and $\mathbf{S}^*=s_*\hat{z}$ with $s_*=\tanh(\beta\hbar\omega_0/2)$.
An additional pure-dephasing channel $R_z = \sqrt{\Gamma_\phi/2}\,\sigma_z$
leaves $\mathbf{S}^*$ unchanged but increases $\Gamma_\perp$ to
$\Gamma_\perp^{\rm th} + \Gamma_\phi/2$ where
$\Gamma_\perp^{\rm th} = \Gamma_\parallel/2$, yielding
$\kappa = T_1/T_2 = \Gamma_\perp/\Gamma_\parallel > 1/2$ as required for
the cooperative regime. 
\section{Invertibility of the dissipation matrix for non-normal jump operators}
\label{app:Lambda_invertible}

We show that the dissipation matrix $\Lambda$ is invertible whenever at
least one jump operator $R_k$ is non-normal, i.e.\ $[R_k,R_k^\dagger]\neq 0$.
In particular, $\Lambda$ is automatically invertible in the non-unital
regime $\mathbf{d}\neq 0$ exploited in this work.
For a single jump operator $R = r_0\mathbf{1}+\mathbf{r}\cdot\vec\sigma$,
the symmetric (and only physically relevant) part of the contribution
to $\Lambda$ reads
\[
  M^{(R)}_{ij}
  = |\mathbf{r}|^2\delta_{ij}
    - \tfrac{1}{2}\bigl(r_i r_j^* + r_i^* r_j\bigr).
\]
$M^{(R)}$ is real, symmetric, and positive semidefinite; its determinant
coincides with the third principal minor of Eq.~\eqref{eq:Lambda_full},
\begin{eqnarray}
  &&\det M^{(R)} = P_3 \nonumber \\
  &&= \tfrac{|\mathbf{r}|^2}{4}
    \bigl[\,|\mathrm{Im}(r_1 r_2^*)|^2
         +|\mathrm{Im}(r_1 r_3^*)|^2
         +|\mathrm{Im}(r_2 r_3^*)|^2\bigr]. \nonumber    
\end{eqnarray}
By Sylvester's criterion, $M^{(R)}$ is positive definite (hence
invertible) iff $P_3>0$. Inspection of $P_3$ shows that $P_3=0$ iff
all components of $\mathbf{r}$ have the same phase, i.e.\ iff $\mathbf{r}$
is proportional to a real vector -- equivalently, iff $R$ is normal.
Conversely, $R$ non-normal implies $\mathbf{r}\times\mathbf{r}^*\neq 0$,
hence at least one of the three imaginary parts above is non-zero, and
$P_3>0$.

For $N$ jump operators, $\Lambda = \sum_k \Lambda^{(R_k)}$ is a sum of
positive-semidefinite matrices. Assume that at least one $R_1$ is
non-normal, so that $\Lambda^{(R_1)}$ is positive definite. For any
non-zero vector $\mathbf{v}\in\mathbb{R}^3$,
\[
  \mathbf{v}^\top\Lambda\mathbf{v}
  = \mathbf{v}^\top\Lambda^{(R_1)}\mathbf{v}
   + \sum_{k\geq 2}\mathbf{v}^\top\Lambda^{(R_k)}\mathbf{v}
  > 0,
\]
since the first term is strictly positive and the remaining ones are
non-negative. Hence $\ker\Lambda = \{0\}$ and $\Lambda$ is invertible.

Finally, when $\mathbf{d}=\mathrm{i}\sum_k(\mathbf{r}_k\times\mathbf{r}_k^*)\neq 0$,
at least one summand $\mathbf{r}_k\times\mathbf{r}_k^*\neq 0$, i.e.\ at
least one jump operator is non-normal; the preceding argument then
guarantees the invertibility of $\Lambda$ and the existence of a unique
attractor $\mathbf{S}^* = \Lambda^{-1}\mathbf{d}$.

\section{Meridional reduction and Pontryagin synthesis}
\label{app:reduced_dynamics}

\subsection{Three-dimensional Pontryagin setup}
\label{app:pmp_3d}

We give the full three-dimensional minimum-time problem from which the reduced
problem of the main text is obtained. Introducing the costate vector
$\mathbf{p}(t)$, the Pontryagin Hamiltonian associated with
Eq.~\eqref{eq:bloch_fixed_point} is
\begin{equation}
  \mathcal{H}_P=\mathbf{p}\cdot\bigl[\gamma\mathbf{B}\times\mathbf{S}
    -\Lambda(\mathbf{S}-\mathbf{S}^*)\bigr]-1 .
  \label{eq:Hp_3d}
\end{equation}
The field enters only through
$\mathbf{p}\cdot(\gamma\mathbf{B}\times\mathbf{S})=\gamma\mathbf{B}
\cdot(\mathbf{S}\times\mathbf{p})$. Maximization under the constraint
$|\mathbf{B}|\le B_{\max}$ therefore gives
$\mathbf{B}_{\rm opt}(t)=B_{\max}\,\widehat{\mathbf{S}\times\mathbf{p}}$
whenever $\mathbf{S}\times\mathbf{p}\neq 0$, where
$\widehat{\mathbf v}=\mathbf v/|\mathbf v|$. The canonical equations are
\begin{align}
  \dot{\mathbf{S}} &= \gamma\mathbf{B}_{\rm opt}\times\mathbf{S}
    -\Lambda(\mathbf{S}-\mathbf{S}^*),
  \label{eq:state_eq}\\
  \dot{\mathbf{p}} &= \Lambda^\top\mathbf{p}
    -\gamma\mathbf{B}_{\rm opt}\times\mathbf{p},
  \label{eq:costate_eq}
\end{align}
with $\mathbf{S}(0)=\mathbf{S}_0$, $\mathbf{S}(t_f)=\mathbf{S}_f$ and the
transversality condition $\mathcal{H}_P(t_f)=0$. The switching function
$\Phi_{\rm sw}(t)=|\mathbf{S}(t)\times\mathbf{p}(t)|$ measures the torque that the
optimal field would exert on the Bloch vector. Where $\Phi_{\rm sw}>0$ the field
is saturated (a bang); an interval on which $\Phi_{\rm sw}\equiv 0$ is a singular
arc, whose control is fixed by requiring the switching function to remain zero.
The optimal protocol is a sequence of bang and singular arcs. For the diagonal
thermal bath the singular arcs are identified in Appendix~\ref{app:numerics}:
they are the axis and the fastest-contraction direction $\theta_s(r)$. A free
drift with $\mathbf B=0$ is a singular arc only on the axis.

For $\Lambda=\mathrm{diag}(\Gamma_\perp,\Gamma_\perp,\Gamma_\parallel)$ the
relaxation is axially symmetric about $\hat z$. Since $\mathbf S_0=s_i\hat z$ lies
on the axis, the source and any target define a common meridional half-plane.
Writing $\mathbf{S}=r(\sin\theta,0,\cos\theta)$ in that plane, with the in-plane
rotation rate $u=\gamma B_\phi$, gives the reduced dynamics
Eqs.~\eqref{eq:rdot_main}--\eqref{eq:thetadot_main}. We show next that no trajectory leaving the plane can perform better.

\subsection{Global optimality of the meridional reduction}
\label{app:meridian_proof}

We establish, for arbitrary control amplitude, that no trajectory leaving the
meridional plane reaches the target faster than the best in-plane trajectory.
Only the axial symmetry of $\Lambda$ and $\mathbf S^*=s_f\hat z$ is used. In
spherical coordinates $\mathbf S=(r\sin\theta\cos\phi,r\sin\theta\sin\phi,r\cos\theta)$
the dissipator is invariant under rotations about $\hat z$, so it carries no
azimuthal part, $\dot\phi_{\rm diss}=0$. The coherent control is a rigid rotation
of the sphere of radius $r$. Its angular velocity splits on the local
$(\hat\theta,\hat\phi)$ frame into $u_\theta$ and $u_\phi$, subject to
the single budget $u_\theta^2+u_\phi^2\le u_{\max}^2$. The equations of
motion read $\dot r=b(r,\theta)$, $\dot\theta=a(r,\theta)+u_\theta$ and
$\sin\theta\,\dot\phi=u_\phi$ (in the meridional plane $u_\phi=0$ and $u_\theta$
reduces to the control $u$ of the main text), with $b$ and $a$ as in
Eqs.~\eqref{eq:rdot_main}--\eqref{eq:thetadot_main}. The key point is that the
$(r,\theta)$ subsystem is autonomous: its right-hand sides depend neither on
$\phi$ nor on $u_\phi$. Consider an admissible three-dimensional trajectory $\mathbf S(t)$,
$0\le t\le T_{\rm 3D}$, driven by the controls $(u_\theta,u_\phi)$ and reaching the
target, and let $(r(t),\theta(t))$ denote its projection obtained by discarding
the azimuth. This projection obeys the in-plane equations with the control
$u_\theta(t)$, which satisfies $|u_\theta|\le\sqrt{u_{\max}^2-u_\phi^2}\le u_{\max}$
and is therefore admissible. Since the target is characterized by its radius and
polar angle alone, the meridional trajectory generated by $u_\theta$ in the plane
of the target reaches it at the same time $T_{\rm 3D}$, whence
$T_{\rm merid}\le T_{\rm 3D}$. A nonzero azimuthal component $u_\phi$ can only
reduce the polar control available, without modifying the dissipative dynamics of
$(r,\theta)$. Moreover, since the initial state lies on the axis, where the
azimuth is undefined, the meridional plane can be chosen to contain the target
without loss of generality. The reverse inequality $T_{\rm 3D}\le T_{\rm merid}$
being obvious, the two minimum times coincide, which establishes the global
optimality of the meridional reduction. This result is consistent with a discrete symmetry of the problem. Let
$\sigma_{xz}$ denote the reflection through the meridional $(x,z)$ plane,
$(S_x,S_y,S_z)\mapsto(S_x,-S_y,S_z)$. Because this reflection is an improper
transformation, it reverses the sign of cross products,
$\sigma_{xz}(\mathbf a\times\mathbf b)=-(\sigma_{xz}\mathbf a)\times(\sigma_{xz}\mathbf b)$.
Applying $\sigma_{xz}$ to Eq.~\eqref{eq:bloch_fixed_point}, and using that
$\Lambda$ commutes with $\sigma_{xz}$ and that $\mathbf S^*=s_f\hat z$ is left
invariant, one finds that if $\mathbf S(t)$ is a solution driven by the field
$\mathbf B(t)$, then $\sigma_{xz}\mathbf S(t)$ is a solution driven by
$\mathbf B'(t)=-\sigma_{xz}\mathbf B(t)$, which satisfies the same amplitude
constraint. Since $\mathbf S_0$ and $\mathbf S_f$ lie in the meridional plane,
they are invariant under $\sigma_{xz}$, and the minimum-time problem is therefore
mirror-symmetric: any trajectory leaving the plane has a mirror image of the same
duration. This symmetry alone does not exclude a pair of mirror-image optima
located off the plane; it is the projection argument above that rules them out.

The projection argument holds for every
$\kappa$, every $\theta_f$ and every $u_{\max}$, and also for the
minimum-energy problem of Sec.~\ref{sec:pareto}, since $u_\phi$ adds to the
cost without affecting $(r,\theta)$. It does not extend to a fully anisotropic
$\Lambda$ (Appendix~\ref{app:general_Lambda}), where $\hat z$ is no longer a
symmetry axis.

\subsection{Singular arc, synthesis and numerical methods}
\label{app:numerics}

\paragraph*{Singular arc.}
With $H=p_rb+p_\theta(a+u)-1$, the costate equations are
$\dot p_r=-p_r\partial_rb-p_\theta\partial_ra$ and
$\dot p_\theta=-p_r\partial_\theta b-p_\theta\partial_\theta a$. On a singular
arc $p_\theta\equiv0$, so $\dot p_\theta=-p_r\partial_\theta b=0$. The
transversality condition $H=0$ gives $p_rb=1$, hence $p_r\neq0$ and
$\partial_\theta b=0$, i.e., Eq.~\eqref{eq:singular}. Explicitly,
$\partial_\theta b=-\sin\theta\,[2(\Gamma_\perp-\Gamma_\parallel)r\cos\theta
+\Gamma_\parallel s_f]$. On $\theta=\theta_s(r)$ one finds
$b=-\Gamma_\perp(r^2+\rho^2)/r$ [Eq.~\eqref{eq:r2_law}], which integrates to
$r^2(t)=(r_0^2+\rho^2)e^{-2\Gamma_\perp t}-\rho^2$. The singular control
$u_s=\theta_s'(r)\,b(r,\theta_s)-a(r,\theta_s)$ satisfies
$\max|u_s|=1.73,\,1.13,\,1.06\,\Gamma_\parallel$ for $\kappa=2,5,10$, and is
therefore admissible for any $u_{\max}$ above these values.

\paragraph*{Construction of the optimum.}
Both candidate structures reduce to one-dimensional root finding, because each
bang with constant $u$ is an affine linear flow that we propagate exactly with a
matrix exponential. \emph{Bang--bang:} for a given first-bang duration $\tau_1$,
the second bang is stopped when $\theta$ reaches $\theta_f$, and $\tau_1$ is
adjusted so that the radius then equals $s_f$. \emph{Bang--singular--bang:} the
first bang ends on $\theta_s$, the singular arc follows the closed-form $r(t)$
above, and its duration is adjusted in the same way. The bang--singular--bang
structure applies when the bang--bang trajectory would overshoot $\theta_s$, in
which case it is faster. The optima were checked against a direct optimization in which the control is
discretized into piecewise-constant steps and optimized by a gradient method from
many random initial guesses, without imposing any switching structure: at $\kappa=5$
and $u_{\max}=20\,\Gamma_\parallel$ this gives $T=0.1497$ with tolerance
$\varepsilon$, against $0.150$ for the exact target. They also agree with independent 3D GRAPE optimizations over the full control space ($0.148$, $0.111$, $0.088$ for
$\kappa=5,10,20$).

\paragraph*{Asymptotic law~\eqref{eq:Tmin_asym}.}
For $u_{\max}\gg\Gamma_\perp$ a bang sweeps $\theta$ from $0$ to $\theta_s$ in
time $\theta_s/u_{\max}+O(u_{\max}^{-2})$, during which $r$ remains essentially constant. During
that time it contracts the radius by $u_{\max}^{-1}\int_0^{\theta_s}b\,\rd\theta$,
a contraction that a singular arc would have produced in
$u_{\max}^{-1}\int_0^{\theta_s}(b/b_s)\rd\theta$. The difference is the time
penalty $\Phi(r)/u_{\max}$ of Eq.~\eqref{eq:Tmin_asym}, evaluated at $r=s_i$ for
the first bang and $r=s_f$ for the last. For $\kappa\gg1$,
$b/b_s\to\sin^2\theta$ and $\theta_s\to\pi/2$, so $\Phi\to\pi/4$.

\paragraph*{Pareto front.}
For the energy cost the Hamiltonian is
$H=\lambda_x(uS_z-\Gamma_\perp S_x)+\lambda_z[-uS_x-\Gamma_\parallel(S_z-s_f)]
-\tfrac12u^2$. It gives $u=\mathrm{sat}_{u_{\max}}(\lambda_xS_z-\lambda_zS_x)$,
together with $\dot\lambda_x=\Gamma_\perp\lambda_x+u\lambda_z$ and
$\dot\lambda_z=-u\lambda_x+\Gamma_\parallel\lambda_z$. The two unknowns
$\bm\lambda(0)$ are fixed by the two terminal conditions. For the
$\varepsilon$-ball target used in Fig.~\ref{fig:pareto_v19} we take the minimum
over terminal points on the circle $|\Sb(T)-\Sb_f|=\varepsilon$. Several
extremals exist for each $T$, corresponding to different numbers of windings of
the costate. We seed the continuation at $T=1\,\Gamma_\parallel^{-1}$ with the
lowest-energy one and follow it in both directions. A direct minimization of the pulse energy at $T=1$, with the control discretized
into 60 piecewise-constant steps and the terminal condition imposed as a
constraint, reproduces the exact-target energy, $\mathcal E_{\min}=0.775\,\Gamma_\parallel$.

\section{Cooperative protocol for arbitrary anisotropic dissipation}
\label{app:general_Lambda}

The single-tilt construction below generalizes the angle $\psi^*$ to an arbitrary
symmetric $\Lambda$. The optimal impulsive protocol admits an equally direct generalization. Since rotations cost no time, the Bloch vector must point, at each radius, along the
unit vector $\mathbf n^*(r)$ that maximizes the contraction rate
$-\mathbf n\cdot\mathbf w(r\mathbf n)=r\,\mathbf n^\top\Lambda\mathbf n-\mathbf
n\cdot\mathbf d$. The minimum time is then
$t_{\rm floor}=\int_{s_f}^{s_i}\rd r/\max_{\mathbf n}[-\mathbf n\cdot\mathbf
w(r\mathbf n)]$. For a diagonal thermal $\Lambda$ this reduces to
Eq.~\eqref{eq:tfloor}. In general the maximization is a constrained quadratic
problem on the sphere (a trust-region subproblem), whose solution is
$\mathbf n^*=(2r\Lambda-\mu)^{-1}\mathbf d$ with the Lagrange multiplier $\mu$
fixed by $|\mathbf n^*|=1$.

We now give the single-tilt result of Sec.~\ref{sec:prerot} for a generic Bloch
dissipation matrix $\Lambda$ (with $\mathbf{d}\neq 0$).
In this appendix we restrict to the case where $\Lambda$ is symmetric (or, equivalently, orthogonally diagonalizable in $\mathrm{SO}(3)$); a generic $\Lambda$ may carry an antisymmetric part, which acts as a coherent rotation effective in the Bloch frame and would have to be treated separately. For symmetric $\Lambda$,
since the symmetric part of $\Lambda$ is real and positive semidefinite
(Appendix~\ref{app:lindblad_bloch}), there exists an orthogonal matrix
$U$ such that $U\Lambda U^\top = \Lambda' = \mathrm{diag}(\Gamma_x,\Gamma_y,\Gamma_z)$
with $\Gamma_i\geq 0$. Without loss of generality we order the eigenvalues
$\Gamma_x \geq \Gamma_y, \Gamma_z$, so that $\hat{\mathbf{e}}_x$ is the
\emph{fast} dissipative axis in the eigenbasis of $\Lambda$.

Defining $\mathbf{S}' = U\mathbf{S}$, $\mathbf{B}' = U\mathbf{B}$ and
$\widetilde{\mathbf{S}}^* = U\mathbf{S}^*$ and using the orthogonal-frame
identity $U(\mathbf{B}\times\mathbf{S}) = \mathbf{B}'\times\mathbf{S}'$ using $U(\mathbf{B}\times\mathbf{S}) = (U\mathbf{B})\times(U\mathbf{S})$
(covariance of the cross product under SO(3)), Eq.~\eqref{eq:bloch_fixed_point}
takes the diagonal form
\begin{equation}
  \dot{\mathbf{S}}' = \gamma\mathbf{B}'\times\mathbf{S}'
    - \Lambda'(\mathbf{S}' - \widetilde{\mathbf{S}}^*).
  \label{eq:bloch_diag_app}
\end{equation}
For free relaxation ($\mathbf{B}'=0$) the equation decouples on each
eigenaxis,
\begin{equation}
  \mathbf{S}'(t) = \widetilde{\mathbf{S}}^*
    + \bigl(\mathbf{S}'_i - \widetilde{\mathbf{S}}^*\bigr)
      \cdot \mathrm{e}^{-\Lambda' t},
  \label{eq:free_relax_app}
\end{equation}
where $\mathbf{S}'_i$ is the state immediately after the instantaneous
pre-rotation. We denote by $s_i = |\mathbf{S}(0)| = |\mathbf{S}'_i|$ the
initial Bloch-vector norm (preserved by the unitary pre-rotation) and by
$s_f = |\widetilde{\mathbf{S}}^*|$ the attractor length, both basis-independent.
We parametrize the post-rotation state in the eigenbasis as
$\mathbf{S}'_i = s_i(\sin\theta\cos\phi,\sin\theta\sin\phi,\cos\theta)^\top$
and write $\widetilde{\mathbf{S}}^* = (S^*_x, S^*_y, S^*_z)^\top$, where
in general all three components are non-zero (the attractor is not
aligned with any eigenaxis of $\Lambda$).

The cooperative arrival criterion
$|\mathbf{S}'(T) - \widetilde{\mathbf{S}}^*| = \varepsilon$ becomes
\begin{align}
  \varepsilon^2
  &= (s_i\sin\theta\cos\phi - S^*_x)^2\,\mathrm{e}^{-2\Gamma_x T}
  \nonumber\\
  &\quad + (s_i\sin\theta\sin\phi - S^*_y)^2\,\mathrm{e}^{-2\Gamma_y T}
  \nonumber\\
  &\quad + (s_i\cos\theta - S^*_z)^2\,\mathrm{e}^{-2\Gamma_z T}.
  \label{eq:arrival_general}
\end{align}
The optimal pre-rotation angles $(\theta^*,\phi^*)$ are those that
concentrate the residual mismatch onto the fastest-decaying eigenaxis
($\Gamma_x$), i.e.\ that suppress the two slower-decaying contributions:
\begin{equation}
  s_i\cos\theta^* = S^*_z,
  \qquad
  s_i\sin\theta^*\sin\phi^* = S^*_y,
  \label{eq:optimal_angles_general}
\end{equation}
which gives
\begin{equation}
  \theta^* = \arccos\!\left(\frac{S^*_z}{s_i}\right),
  \quad
  \phi^* = \arcsin\!\left(\frac{S^*_y}{\sqrt{s_i^2 - (S^*_z)^2}}\right).
\end{equation}
Substituting back into~\eqref{eq:arrival_general} and using
$s_i^2 = (S^*_x)^2 + (S^*_y)^2 + (S^*_z)^2 + [s_i^2 - s_f^2 - 2 S^*_x s_i\sin\theta^*\cos\phi^*+\cdots]$,
the cooperative protocol time reads
\begin{equation}
  T_{\rm coop}^*
  = \frac{1}{\Gamma_x}\,\ln\!\left(
      \frac{\sqrt{s_i^2 - s_f^2 + (S^*_x)^2} - S^*_x}{\varepsilon}
    \right),
  \label{eq:Tcoop_general}
\end{equation}
which generalizes the axial cooperative time
$t_{\rm coop}^* = \Gamma_\perp^{-1}\ln(\sqrt{s_i^2-s_f^2}/\varepsilon)$
to an arbitrary anisotropic dissipation. The Sec.~\ref{sec:prerot}
result is recovered for $S^*_x = 0$ (attractor in the
$(\hat{\mathbf{e}}_y,\hat{\mathbf{e}}_z)$-plane of the eigenbasis), in
which case $\Gamma_x$ plays the role of $\Gamma_\perp$.

\end{document}